\documentclass[12pt]{article}

\usepackage[utf8]{inputenc}
\usepackage[T1]{fontenc}
\usepackage[english]{babel}

\usepackage{amsmath,amssymb,amsthm,mathtools}
\usepackage{bm}
\usepackage{multicol}
\usepackage{geometry}
\usepackage{graphicx}
\usepackage{subcaption}

\usepackage{xcolor}
\usepackage{float}
\usepackage[section]{placeins}

\usepackage{csquotes}
\usepackage{hyperref}
\usepackage[nameinlink,capitalize]{cleveref}

\usepackage{siunitx}
\usepackage{icomma}
\usepackage{booktabs,tabularx}

\usepackage[backend=biber,style=phys,sorting=none,
 articletitle=false,chaptertitle=false,
 biblabel=brackets,maxbibnames=6,giveninits=true,
 doi=false,isbn=false,url=false,eprint=true]{biblatex}
\title{\textbf{A Descriptor Surrogate for Kerr Shadow Contours}}

\author{Arseny Pantsialei\thanks{e-mail: wselend@gmail.com} \\
\emph{Institute of Physics, Maria Curie-Sk\l{}odowska University, 20-031 Lublin, Poland}}
\date{}

\begin{document}

\maketitle

\begin{abstract}
We consider the Kerr black hole shadow contour as a geometric object whose essential shape can be captured by a small set of physically transparent descriptors. Starting from exact critical-curve calculations, 
we map each contour to a centered radial profile and retain five quantities: the mean radius, the horizontal centroid shift, and three low-order harmonic coefficients. In this way, the aim is to obtain a compact 
contour-level representation suitable for repeated evaluation and systematic comparison. We then build a boundary-aware surrogate 
over spin and inclination that recovers the correct circular limits near the Schwarzschild and polar boundaries and includes a simple positivity safeguard for the reconstructed radial profile. On a held-out bulk 
domain with inclinations $i \ge 5^\circ$, the production surrogate yields median errors of 0.522\% in contour area, 0.261\% in equivalent diameter, and 0.954\% in the 95th-percentile radial contour mismatch; the 
corresponding 95th-percentile errors are 1.100\%, 0.551\%, and 3.436\%, with no negative-radius samples. The exact descriptor reconstruction is already accurate at the sub-percent level for area and equivalent 
diameter, which shows that the descriptor layer is a meaningful compression of the contour geometry. We also show that contours with nearly identical equivalent diameter can 
remain visibly distinct, so the retained descriptors encode geometric information beyond a single size observable. The resulting model provides a compact and interpretable contour-level surrogate for parameter 
scans and repeated contour comparisons in the Kerr problem.
\end{abstract}

\section{Introduction}

Black-hole shadow observations at horizon-scale resolution have made the geometry of the shadow contour a physically meaningful object in its own right. It is no longer enough to regard the contour as a mere visual 
outline. In present applications it already enters model comparison, parameter inference, and the interpretation of increasingly refined observations 
\cite{EHTC2024_SgrA_Polarization_VIII,EHTC2024_M87_Persistent_I,EHTC2025_M87_Persistent_II,Tiede2022_ngEHT_PhotonRing,Walia_2025,Wielgus2024_SgrA_InternalFaraday,Vagnozzi:2023SgrA_CQG,Khodadi:2024MimeticSR}. At the 
same time, much of the existing literature still compresses the shadow to one or a few scalar observables, for example a characteristic size, a mean radius, a circularity measure, or a distortion parameter 
\cite{PhysRevD.80.024042,PhysRevD.100.044057,KumarGhosh2020,Sui2023_Accel_KerrNewman_ArealRadius,AgurtoSepulveda2024_AlphaCorrectedShadows,Kuang:2024ugn,Ban2025_EPJC_QGShadows,PhysRevD.108.044008}. Such quantities 
are certainly useful, but they leave open a more elementary geometric question: to what extent can the information contained in the full Kerr shadow contour be retained in a compact low-dimensional form?

This question should be distinguished from the construction of a fast approximation for a single scalar observable. A one-parameter surrogate may be convenient in practice, yet it says nothing about the geometry of 
the contour itself. A size measure such as the equivalent diameter $D_{\rm eq}$ can characterize an overall scale, but it cannot by itself record the displacement of the centroid, the leading asymmetries, or the 
first deviations from circularity that change across the Kerr parameter space. It follows that two Kerr shadows may have nearly the same scalar size and still differ appreciably in shape. The contour, therefore, 
is not merely a more detailed version of a scalar observable; it is a different geometric object.

The purpose of the present work is to isolate and test a compact set of low-order global descriptors for the Kerr shadow contour, obtained directly from the exact critical curve, and to show that these descriptors 
already suffice to reconstruct the contour accurately over the physically relevant bulk of parameter space. Our construction proceeds in a natural sequence. We begin with an exact deterministic mapping from Kerr 
geodesics to an ordered shadow contour. We then pass to a descriptor representation based on the radial profile of the contour after centering. Finally, we construct a boundary-aware surrogate for the resulting 
descriptor fields over the plane of spin and inclination, $(a_*,i)$. The role of this surrogate is not to replace exact contour generation where ultimate precision is required. Its role is more modest and, at the 
same time, more transparent: it provides a compact contour-level representation that is easy to evaluate, computationally inexpensive, and directly interpretable in scans and comparisons.

The primary aim of this paper is geometric rather than inferential. Our task here is more limited and, for that reason, more transparent. We construct and test a compact, interpretable representation of the exact 
Kerr critical contour. Only after this compression step 
do we introduce the surrogate layer, whose purpose is to enable rapid contour-level scans and comparisons, where repeatedly storing or evaluating dense contours is less convenient than working with a small 
descriptor vector.

In this sense, the main statement of the paper is geometric before it is computational. The essential point is that the Kerr shadow possesses a low-order descriptor basis with substantial intrinsic reconstruction 
power even before any emulation is introduced. The surrogate is built only afterward, and it acts on this descriptor layer rather than on the raw contour. This distinction is important. The surrogate is not 
compensating for an inadequate representation; it approximates an object that already captures physically meaningful information beyond any single size observable. One may therefore view the present construction as 
a complement to existing characterizations based on diameter, circularity, asymmetry, and related integrated quantities 
\cite{PhysRevD.80.024042,PhysRevD.100.044057,KumarGhosh2020,Sui2023_Accel_KerrNewman_ArealRadius,AgurtoSepulveda2024_AlphaCorrectedShadows,Kuang:2024ugn,Ban2025_EPJC_QGShadows,PhysRevD.108.044008}. Those quantities 
summarize the shadow. Here we seek instead a compact representation of the contour itself.

From this point of view, the contributions of the paper are straightforward. We first define a minimal low-order descriptor basis for the exact Kerr contour and examine its reconstruction power independently of any 
surrogate model. We then build a practical production-level surrogate for these descriptors on a dense grid in $(a_*,i)$ and test it on held-out parameter points. Finally, we show explicitly that nearly equal 
values of $D_{\rm eq}$ do not imply the same Kerr shadow shape. The contour-level descriptors therefore retain geometric information that is invisible to a single size measure. This makes the construction useful 
as a compact geometric language for Kerr-shadow shape in parameter scans and contour-level comparisons, and it may also serve as a convenient starting point for future forward-model developments 
\cite{Tiede2022_ngEHT_PhotonRing,Walia_2025,EHTC2025_M87_Persistent_II}.

\section{Exact Kerr contour pipeline}
\label{sec:exact_pipeline}

Our construction begins with the exact critical curve of the Kerr geometry. Throughout, we use the standard conventions for Kerr geodesics and for celestial coordinates on the observer screen 
\cite{LandauLifshitzFields,PhysRevD.80.024042}. We work in geometric units, setting $G=c=1$, and unless stated otherwise we also set $M=1$. The restoration of physical units is then entirely straightforward, 
since the screen coordinates scale as $(\alpha,\beta)\mapsto M(\alpha,\beta)$. As a result, the enclosed area scales as $A\mapsto M^2A$, and the equivalent diameter scales linearly, $D_{\rm eq}\mapsto M D_{\rm eq}$. Thus the geometry of the contour is determined once and for all in dimensionless form, while the black-hole mass simply fixes the overall angular scale.

For each pair $(a_*,i)$, the shadow contour is obtained deterministically from the family of unstable spherical photon orbits. This is the familiar critical-curve construction used in Kerr-shadow studies:
 the boundary of the shadow is formed by those null geodesics that asymptotically separate capture from escape, and these geodesics are precisely associated with spherical photon motion 
 \cite{PhysRevD.80.024042,PhysRevD.100.044057}. In this sense, the contour is not introduced phenomenologically, but follows directly from the geodesic structure of the spacetime.

The null geodesics admit three conserved quantities, and it is convenient to express them through the usual impact parameters
\begin{equation}
\xi \equiv \frac{L_z}{E},
\qquad
\eta \equiv \frac{Q}{E^2},
\end{equation}
where $E$ is the conserved energy of the photon, $L_z$ is the component of angular momentum along the symmetry axis, and $Q$ is the Carter constant. Physically, $\xi$ controls the azimuthal part of the motion, 
while $\eta$ measures the part of the motion that cannot be reduced to the equatorial plane alone. Introducing
\begin{equation}
\Delta(r)=r^2-2r+a_*^2,
\end{equation}
we can write the radial potential for null geodesics in the form
\begin{equation}
R(r)=\bigl(r^2+a_*^2-a_*\xi\bigr)^2
-\Delta\Bigl[\eta+(\xi-a_*)^2\Bigr].
\end{equation}
This function $R(r)$ governs the radial motion: where it vanishes, the radial velocity vanishes, and where in addition its derivative vanishes, the orbit remains at fixed radius. It is therefore natural that 
spherical photon orbits are characterized by the conditions

\begin{equation}
R(r)=0,
\qquad
R'(r)=0.
\end{equation}
Solving these two equations for the impact parameters gives
\begin{equation}
\xi(r)=
\frac{r^2(3-r)-a_*^2(r+1)}{a_*(r-1)},
\label{eq:xi_of_r}
\end{equation}
and
\begin{equation}
\eta(r)=
\frac{r^3\!\left(4a_*^2-r(r-3)^2\right)}{a_*^2(r-1)^2}.
\label{eq:eta_of_r}
\end{equation}
Thus each admissible spherical radius $r$ determines a definite pair $(\xi,\eta)$, and with it a definite point on the shadow boundary.

To pass from the constants of motion to the observed image, we use the standard celestial-coordinate map for an observer at infinity. The horizontal screen coordinate is
\begin{equation}
\alpha(r)=-\frac{\xi(r)}{\sin i},
\label{eq:alpha_of_r}
\end{equation}
while the vertical coordinate is
\begin{equation}
\beta_{\pm}(r)=
\pm\sqrt{\eta(r)+a_*^2\cos^2 i-\xi(r)^2\cot^2 i}.
\label{eq:beta_of_r}
\end{equation}
The meaning of these expressions is simple. The parameter $\alpha$ measures the apparent displacement of the ray along the projected equatorial direction, whereas $\beta$ gives the displacement in the 
orthogonal direction on the observer screen. The two signs in $\beta_\pm$ correspond to the upper and lower branches of the critical curve. As the spherical-orbit radius varies over its allowed interval, 
these branches together trace the full shadow contour.

The allowed interval itself is bounded by the prograde and retrograde equatorial photon orbits. Their radii are
\begin{equation}
r_-=
2\left[1+\cos\!\left(\frac{2}{3}\arccos(-a_*)\right)\right],
\label{eq:rminus}
\end{equation}
and
\begin{equation}
r_+=
2\left[1+\cos\!\left(\frac{2}{3}\arccos(+a_*)\right)\right].
\label{eq:rplus}
\end{equation}
These two limiting radii mark the extremal spherical photon orbits that contribute to the contour. In the Schwarzschild limit, $a_*=0$, they coincide at $r_\pm=3$, which is exactly what one expects, since in the 
nonrotating case all unstable photon orbits collapse to the single photon sphere at $r=3$.

For numerical work, we sample a monotone grid $\{r_j\}_{j=1}^{N_r}$ on the interval $[r_-,r_+]$. From this grid we construct the contour in the natural way: first we follow the upper branch through the points 
$\bigl(\alpha(r_j),\beta_+(r_j)\bigr)$, and then we return along the lower branch in reverse order through the points $\bigl(\alpha(r_j),\beta_-(r_j)\bigr)$ for $j=N_r,\dots,1$. In this way we obtain an ordered 
closed polygon,
\begin{equation}
\mathcal{C}
=
\{(\alpha_k,\beta_k)\}_{k=1}^{N},
\qquad
(\alpha_{N+1},\beta_{N+1})\equiv(\alpha_1,\beta_1),
\end{equation}
which serves as a deterministic approximation to the exact critical curve. The important point is that no ambiguity enters this construction: once $(a_*,i)$ is fixed, the contour is fixed as well, and the resulting
 polygonal curve provides the starting object for the descriptor analysis that follows.

Once the contour points have been ordered, the basic geometric observables can be obtained directly from the polygon itself. The enclosed area is computed by the standard shoelace formula,
\begin{equation}
A=\frac{1}{2}\left|
\sum_{k=1}^{N}\left(\alpha_k\beta_{k+1}-\alpha_{k+1}\beta_k\right)
\right|,
\label{eq:shoelace_area}
\end{equation}
which is simply the discrete expression for the area enclosed by the shadow boundary. From this area we define the equivalent diameter,
\begin{equation}
D_{\rm eq}\equiv 2\sqrt{\frac{A}{\pi}},
\label{eq:Deq_def}
\end{equation}
that is, the diameter of the circle having the same area as the shadow. We retain this quantity in the present work as a derived measure of size, since it provides a convenient global scale. At the same time, it 
should be clear that such a quantity cannot be the main geometric object. By construction, the equivalent diameter, like related one-number summaries used in the shadow literature, compresses the entire contour 
to a single scalar and therefore discards most of the information carried by the boundary itself \cite{PhysRevD.100.044057,KumarGhosh2020}.

A separate remark is needed for the axial limit $i\to 0$. The celestial-coordinate expressions given above are written for generic observer inclination and become formally singular when $\sin i\to 0$ and 
$\cot i\to\infty$. This singularity is only apparent. Physically, the shadow seen by an observer on the axis must be circular. To make this manifest, it is convenient to introduce the sky-plane radius
\begin{equation}
\rho^2\equiv \alpha^2+\beta^2
=
\eta+\xi^2+a_*^2\cos^2 i.
\end{equation}
This quantity remains regular in the axial limit and makes the geometry transparent. For an exactly polar observer, regularity requires $\xi\to 0$, so the shadow radius is determined by the unique spherical 
photon orbit for which
\begin{equation}
\xi(r_0)=0,
\qquad
\rho_{\rm pole}(a_*)=\sqrt{\eta(r_0)+a_*^2}.
\label{eq:rho_pole}
\end{equation}
Using Eq.~\eqref{eq:xi_of_r}, the condition $\xi(r_0)=0$ becomes
\begin{equation}
r_0^3-3r_0^2+a_*^2 r_0+a_*^2=0,
\label{eq:r0_cubic}
\end{equation}
and the physically relevant root is selected by continuity with the Schwarzschild case, for which $r_0(0)=3$. The polar shadow is therefore a circle with area and equivalent diameter
\begin{equation}
A_{\rm pole}=\pi \rho_{\rm pole}^2,
\qquad
D_{{\rm eq},\,{\rm pole}}=2\rho_{\rm pole}.
\end{equation}
Thus even in the axial case the same construction leads to a completely regular geometric description, once one uses the appropriate invariant radius on the observer screen.

The same exact contour data are then used for the extraction of descriptors. This point is conceptually important for everything that follows. The surrogate introduced later is not trained on an artificial image 
model, nor on a black-box latent representation with obscure geometric meaning. It is trained on deterministic contour data obtained directly from the exact Kerr geometry. In this sense, the approximation enters 
only after the geometric object itself has been defined in a precise and physically transparent way.

\section{Low-order descriptor construction}
\label{sec:descriptor_construction}

It is useful to make clear what kind of compression we are actually seeking. The exact Kerr critical curve is already a complete geometric object, but completeness 
by itself is not yet understanding. If one keeps the contour only as a dense ordered set of points, then all information is present, yet little is organized. What we need instead is a representation in 
which the physically distinct ingredients of the shadow are separated: the overall size, the trivial displacement on the observer screen, and the genuinely shape-carrying deviations from circularity.

This separation is natural from the geometric point of view. A smooth closed contour in the image plane may be described either by its raw Cartesian coordinates or, once a suitable center has been chosen, 
by its distance from that center as a function of angle. For the Kerr shadow the second description is especially convenient. The contour is smooth, nearly circular over much of parameter space, and its 
departures from circularity are global rather than erratic. One therefore expects that the leading geometry should be captured not by many unrelated local parameters, but by a small number of low-order modes 
with direct geometric meaning.

Our aim in this section is precisely to pass from the exact contour to such a minimal description. We do not seek a formal spectral expansion for its own sake, nor an arbitrarily accurate decomposition with 
many coefficients. What matters here is more physical: we want the smallest basis that still retains the dominant global structure of the Kerr shadow. Once this is achieved, the contour is no longer represented 
as a mere collection of points, but as a compact geometric object whose components can be interpreted directly.

\subsection{Centered radial representation}

We now pass from the ordered exact contour
\(
\mathcal{C}=\{(\alpha_k,\beta_k)\}_{k=1}^{N}
\)
to a representation in which the trivial overall displacement of the shadow has been removed. This is naturally done by introducing the contour-centered frame. For that purpose, we first compute the signed polygon 
area,
\begin{equation}
A_{\rm s}
=
\frac{1}{2}
\sum_{k=1}^{N}
\left(
\alpha_k \beta_{k+1}
-
\alpha_{k+1}\beta_k
\right),
\qquad
(\alpha_{N+1},\beta_{N+1})\equiv (\alpha_1,\beta_1),
\end{equation}
from which the centroid of the polygon follows in the standard way:
\begin{equation}
\alpha_c
=
\frac{1}{6A_{\rm s}}
\sum_{k=1}^{N}
(\alpha_k+\alpha_{k+1})
\left(
\alpha_k \beta_{k+1}
-
\alpha_{k+1}\beta_k
\right),
\label{eq:alpha_centroid}
\end{equation}
and
\begin{equation}
\beta_c
=
\frac{1}{6A_{\rm s}}
\sum_{k=1}^{N}
(\beta_k+\beta_{k+1})
\left(
\alpha_k \beta_{k+1}
-
\alpha_{k+1}\beta_k
\right).
\label{eq:beta_centroid}
\end{equation}
The meaning of this step is simple. A shift of the entire contour on the observer screen does not change its intrinsic shape, and it is therefore natural to separate this trivial displacement from the geometry 
we actually wish to describe. We accordingly introduce the centered coordinates
\begin{equation}
\tilde\alpha_k=\alpha_k-\alpha_c,
\qquad
\tilde\beta_k=\beta_k-\beta_c.
\end{equation}

Once the contour has been recentered in this way, its shape can be described by a radial profile with respect to the centroid. We write
\begin{equation}
\tilde\alpha = R(\psi)\cos\psi,
\qquad
\tilde\beta = R(\psi)\sin\psi,
\label{eq:centered_polar_profile}
\end{equation}
where
\begin{equation}
\psi = \operatorname{atan2}(\tilde\beta,\tilde\alpha)
\end{equation}
is the polar angle around the centroid. Thus the entire contour is encoded in the single function $R(\psi)$, which gives the distance from the centroid to the boundary in each angular direction. This is already 
a considerable simplification: instead of treating the contour as an unordered cloud of points in the screen plane, we represent it as a one-dimensional geometric object, namely a radial function on the circle.

For the exact Kerr contours considered here, this representation is particularly natural. The boundary is smooth and star-shaped with respect to the centroid, so every ray emitted from the centroid at fixed 
angle $\psi$ intersects the contour exactly once. It follows that $R(\psi)$ is single-valued. In other words, no information is lost in passing to the centered radial description. We have merely replaced the 
original planar contour by a more economical form in which the intrinsic geometry of the shadow becomes easier to analyze.

For numerical purposes, the exact contour is resampled on a uniform angular grid,
\begin{equation}
\psi_m=\frac{2\pi m}{N_\psi},
\qquad
m=0,1,\dots,N_\psi-1,
\end{equation}
and we denote the corresponding sampled radial profile by
\begin{equation}
R_m \equiv R(\psi_m).
\end{equation}
The first quantity of interest is the mean radius,
\begin{equation}
\bar R
=
\frac{1}{2\pi}\int_{0}^{2\pi} R(\psi)\,d\psi
\;\approx\;
\frac{1}{N_\psi}\sum_{m=0}^{N_\psi-1} R_m.
\label{eq:Rbar_def}
\end{equation}
This quantity fixes the overall scale of the centered contour. What remains after dividing by this scale is the genuine angular structure of the shadow. It is therefore natural to introduce the 
dimensionless fluctuation
\begin{equation}
f(\psi)\equiv \frac{R(\psi)}{\bar R}-1,
\qquad
\frac{1}{2\pi}\int_{0}^{2\pi} f(\psi)\,d\psi = 0.
\label{eq:fpsi_def}
\end{equation}
Thus $f(\psi)$ measures the deviation of the contour from a circle of radius $\bar R$, and by construction it has zero mean.

Once the radial profile has been written in this form, the next step suggests itself. Since $f(\psi)$ is a periodic function on the circle, its angular structure can be resolved into harmonics. 
We therefore introduce the Fourier coefficients
\begin{equation}
d_n
=
\frac{1}{2\pi}\int_{0}^{2\pi} f(\psi)e^{-in\psi}\,d\psi
\;\approx\;
\frac{1}{N_\psi}\sum_{m=0}^{N_\psi-1} f(\psi_m)e^{-in\psi_m},
\qquad n\ge 1.
\label{eq:dn_def}
\end{equation}
These coefficients provide a natural global description of the contour: each harmonic isolates a definite angular deformation mode.

In the Kerr problem considered here, the situation simplifies further. After centering the contour, and using the standard reflection symmetry
\(
\beta\mapsto -\beta
\),
the dominant low-order information lies in the real cosine sector. In particular, the vertical centroid satisfies
\(
\beta_c=0
\)
up to numerical precision, so the leading nontrivial structure is already well captured by the real coefficients
\begin{equation}
c_n \equiv \Re d_n,
\qquad n=2,3,4.
\label{eq:cn_from_dn}
\end{equation}
Equivalently, one may write
\begin{equation}
c_n
=
\frac{1}{2\pi \bar R}
\int_{0}^{2\pi}
\bigl(R(\psi)-\bar R\bigr)\cos(n\psi)\,d\psi
\;\approx\;
\frac{1}{N_\psi \bar R}
\sum_{m=0}^{N_\psi-1}
\bigl(R_m-\bar R\bigr)\cos(n\psi_m).
\label{eq:cn_real_def}
\end{equation}
This form makes the meaning particularly transparent: the coefficients $c_n$ measure the projection of the contour deformation onto the lowest cosine harmonics.

We are thus led to approximate the centered radial profile by the low-order expansion
\begin{equation}
R(\psi)
\approx
\bar R
\Bigl[
1
+
2\bigl(
c_2\cos(2\psi)
+
c_3\cos(3\psi)
+
c_4\cos(4\psi)
\bigr)
\Bigr].
\label{eq:low_order_radial}
\end{equation}
In other words, the exact contour is replaced by a circle of radius $\bar R$, corrected by the first few global angular deformations. The reconstructed curve on the observer screen then takes the form
\begin{equation}
\alpha_{\rm rec}(\psi)=\alpha_c + R(\psi)\cos\psi,
\qquad
\beta_{\rm rec}(\psi)=\beta_c + R(\psi)\sin\psi.
\label{eq:reconstructed_contour}
\end{equation}
Thus the passage back to screen coordinates is immediate once the radial profile is known.

It is now clear which quantities should be retained as the basic descriptors. We adopt the set
\begin{equation}
\mathcal{D}=\{\bar R,\alpha_c,c_2,c_3,c_4\}.
\label{eq:descriptor_set}
\end{equation}
Its interpretation is straightforward. The mean radius $\bar R$ fixes the overall size of the contour. The centroid coordinate $\alpha_c$ describes its horizontal displacement on the screen. 
The coefficient $c_2$ gives the leading quadrupolar deviation from circularity. The coefficient $c_3$ captures the first fore-aft asymmetry associated with rotation. Finally, $c_4$ supplies the 
next global correction to the contour shape. In this way, the shadow is described not by a single integrated number, but by a small set of geometrically transparent quantities that still retain the 
leading structure of the full boundary.

\subsection{Why this is a minimal useful basis}

A natural question is why the descriptor set
\(
\{\bar R,\alpha_c,c_2,c_3,c_4\}
\)
should be regarded as a minimal useful basis. The point is not to pursue an arbitrarily accurate spectral expansion of the contour, but to isolate the smallest low-order set that already captures the dominant 
global geometry of the Kerr shadow. Once the contour has been centered, the $n=1$ harmonic disappears by construction, since it would merely describe a residual translation of the boundary rather than an 
intrinsic deformation. The first harmonic that carries genuine shape information is therefore $n=2$.

From this one can already see why a smaller basis is insufficient. If one were to retain only $\bar R$ and $\alpha_c$, then contours that are visibly different would still be compressed into nearly the same 
representation. The overall size and the horizontal displacement would be preserved, but the leading deformation away from circularity would be lost. On the other hand, keeping only a single additional harmonic 
is generally not enough to reproduce the characteristic asymmetry of the Kerr shadow throughout the bulk of parameter space. The contour is shaped not only by an overall quadrupolar deformation, but also by the 
fore-aft asymmetry induced by rotation and by the next global correction that refines this structure.

For this reason, the set
\(
\{\bar R,\alpha_c,c_2,c_3,c_4\}
\)
is the smallest basis we found that still reproduces the leading contour variations in a stable and physically transparent way. A smaller choice loses essential geometric information, while a substantially larger 
one would diminish the gain of compression and make the individual roles of the descriptors less clear. Equation~\eqref{eq:low_order_radial} should therefore be viewed not as a maximal expansion, but as a 
practical compromise between compactness, interpretability, and reconstruction accuracy.

One can make this truncation statement quantitative by examining the harmonic power of the centered fluctuation $f(\psi)$ introduced above. With the coefficients $d_n$ from Eq.~\eqref{eq:dn_def}, we define
\begin{equation}
P_n \equiv |d_n|^2,
\qquad
C_N \equiv \frac{\sum_{n=2}^{N} P_n}{\sum_{n=2}^{N_{\max}} P_n},
\qquad
T_{\ge 5} \equiv 1-C_4,
\label{eq:cumulative_power_tail}
\end{equation}
where $N_{\max}$ denotes the highest resolved harmonic on the sampled angular grid. We also monitor the absolute amplitude of the discarded tail through
\begin{equation}
\varepsilon_{\ge 5}^{\rm rms}
\equiv
100\sqrt{2\sum_{n=5}^{N_{\max}} P_n},
\label{eq:tail_rms_amplitude}
\end{equation}
which measures the rms contribution of the omitted modes relative to $\bar R$. Evaluated on the exact $60\times 91$ grid used throughout the paper, the retained modes $n=2,3,4$ carry $96.6\%$ of the aggregate 
resolved noncircular power over the full domain and $99.0\%$ over the bulk domain $i\ge 5^\circ$. The pointwise median value in the bulk domain is $C_4=97.1\%$. In the same bulk regime, the omitted tail has 
$\varepsilon_{\ge 5}^{\rm rms}=0.042\%$ at the median and $0.231\%$ at the $p95$ level. Thus the $n\ge 5$ sector lies below the exact-level truncation floor quantified later in Table~\ref{tab:exact_recon_summary}, 
so extending the reconstruction beyond $c_4$ changes the contour only marginally in the regime where the model is intended to operate.

\section{Exact descriptor-level reconstruction}
\label{sec:exact_reconstruction}

Before introducing any surrogate, we must first understand what is already contained in the descriptor layer itself. The primary question at this stage is not yet whether the descriptor fields can be interpolated 
accurately across parameter space. The more basic question is whether the exact Kerr contour can already be compressed into the low-order basis
\begin{equation}
\mathcal D \equiv \{\bar R,\alpha_c,c_2,c_3,c_4\}
\label{eq:descriptor_set_exact_recon}
\end{equation}
without losing the dominant geometry of the boundary.

This distinction is conceptually important. If the descriptors extracted from the exact critical curve were themselves unable to reconstruct the contour with sufficient fidelity, then even a highly successful 
surrogate fit for these descriptors would carry little geometric significance. One would merely be approximating an inadequate representation with great efficiency. The exact-level reconstruction test is 
therefore needed to separate two logically different issues. The first is the intrinsic compression power of the descriptor basis itself. The second is the additional approximation error introduced only later, 
when this already-defined descriptor layer is replaced by a surrogate model.

\begin{figure*}[t]
  \centering
  \includegraphics[width=\textwidth]{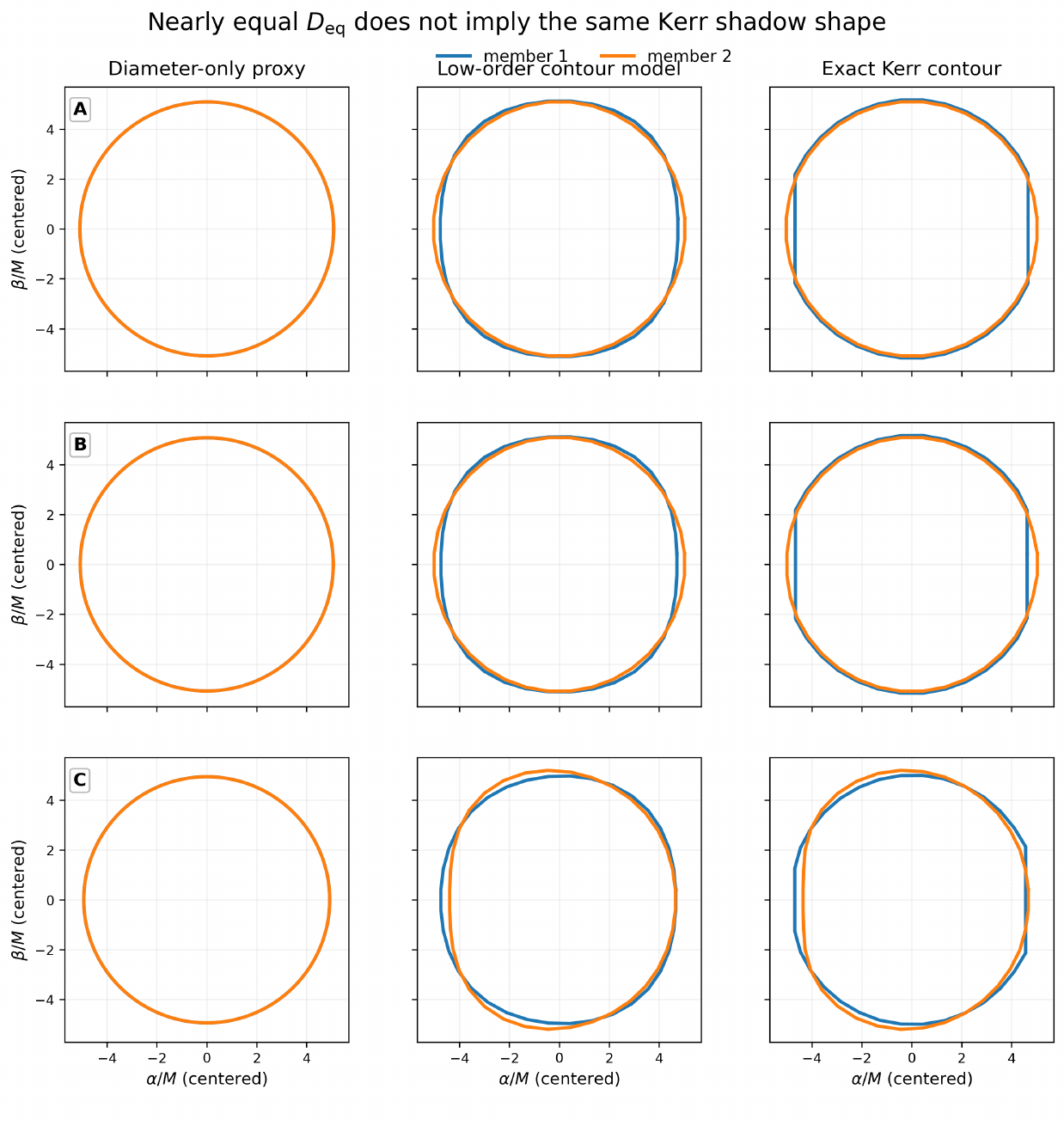}
  \caption{
   Nearly equal equivalent diameter does not imply the same Kerr shadow shape.
Each row shows a pair of Kerr shadows with nearly identical $D_{\rm eq}$. Left: the diameter-only
proxy collapses each pair onto nearly the same circle. Middle: the low-order descriptor reconstruction
resolves the contour-level difference. Right: the corresponding exact Kerr contours confirm that this
difference is physical rather than a surrogate artifact. This diagnostic example illustrates that the descriptor set retains geometric information beyond a single size observable.
  }
  \label{fig:deq_shape}
\end{figure*}

\subsection{Reconstruction protocol and error measures}
\label{subsec:exact_recon_protocol}

For each exact grid point $(a_\ast,i)$, we extract the descriptor set $\mathcal D(a_\ast,i)$ from the exact Kerr contour and then reconstruct the corresponding low-order radial profile from the 
descriptor expansion. In the present notation this reconstructed profile is
\begin{equation}
R_{\rm rec}(\psi)
=
\bar R
\Big[
1 + 2\bigl(
c_2 \cos(2\psi)
+
c_3 \cos(3\psi)
+
c_4 \cos(4\psi)
\bigr)
\Big].
\label{eq:Rrec_exact}
\end{equation}
Thus the exact contour is replaced by its projection onto the low-order descriptor basis, with $\bar R$ setting the overall scale and the coefficients $c_2$, $c_3$, and $c_4$ carrying the leading global 
deformations. The corresponding reconstructed contour on the observer screen is then written as
\begin{equation}
\alpha_{\rm rec}(\psi)=\alpha_c+R_{\rm rec}(\psi)\cos\psi,
\qquad
\beta_{\rm rec}(\psi)=\beta_c+R_{\rm rec}(\psi)\sin\psi,
\label{eq:contour_rec_exact}
\end{equation}
where in the Kerr setup considered here one has $\beta_c=0$ up to numerical precision.

To compare the reconstructed contour with the exact one, both are evaluated on the same uniform angular grid,
\begin{equation}
\psi_m = \frac{2\pi m}{N_\psi},
\qquad
m=0,1,\dots,N_\psi-1.
\label{eq:psi_grid_exact_recon}
\end{equation}
We denote the exact and reconstructed radii on this grid by
\begin{equation}
R^{\rm ex}_m \equiv R_{\rm ex}(\psi_m),
\qquad
R^{\rm rec}_m \equiv R_{\rm rec}(\psi_m).
\label{eq:R_exact_and_rec_samples}
\end{equation}
This places the comparison in the most direct possible form: at each angular direction we compare the true radial distance to the one predicted by the low-order basis. In this way the question is no 
longer qualitative. One can test explicitly how much of the exact contour survives the compression to a few global descriptors.

From the reconstructed contour we compute the enclosed area $A_{\rm rec}$ by the same shoelace construction used for the exact contour, and from it the reconstructed equivalent diameter,
\begin{equation}
D_{{\rm eq},\,{\rm rec}}
=
2\sqrt{\frac{A_{\rm rec}}{\pi}}.
\label{eq:Deq_rec_exact}
\end{equation}
The corresponding exact quantities are denoted by $A_{\rm ex}$ and $D_{{\rm eq},\,{\rm ex}}$. It is worth stressing that these are not independent observables added from outside. They are derived 
directly from the reconstructed contour itself, and therefore they test whether the low-order basis preserves not only local shape but also the integrated geometry of the boundary.

To monitor the reconstruction quality, we use three complementary diagnostics. The first is the relative area error,
\begin{equation}
\Delta A
\equiv
100\,\frac{|A_{\rm rec}-A_{\rm ex}|}{A_{\rm ex}} .
\label{eq:deltaA_exact}
\end{equation}
This measures how accurately the compressed contour reproduces the total enclosed region on the observer screen. Since the area is one of the most global characteristics of the shadow, agreement here 
tells us whether the overall scale and broad geometry have been retained.

The second diagnostic is the relative error in the equivalent diameter,
\begin{equation}
\Delta D_{\rm eq}
\equiv
100\,\frac{|D_{{\rm eq},\,{\rm rec}}-D_{{\rm eq},\,{\rm ex}}|}
{D_{{\rm eq},\,{\rm ex}}} .
\label{eq:deltaDeq_exact}
\end{equation}
Although the equivalent diameter is only a derived scalar summary, it remains useful because it translates the area into the familiar language of an effective size. In this sense, $\Delta D_{\rm eq}$ tells 
us how much of the standard size characterization is preserved by the descriptor-level reconstruction.

The third diagnostic probes the contour more directly. At each sampled angle we define the radial mismatch
\begin{equation}
\delta r_m
\equiv
100\,\frac{|R^{\rm rec}_m-R^{\rm ex}_m|}{R^{\rm ex}_m},
\qquad
m=0,1,\dots,N_\psi-1,
\label{eq:delta_r_m_exact}
\end{equation}
and summarize this angular distribution by its empirical 95th percentile,
\begin{equation}
\Delta r_{95}
\equiv
Q_{0.95}\!\left(\{\delta r_m\}_{m=0}^{N_\psi-1}\right).
\label{eq:delta_r95_exact}
\end{equation}
This choice is natural. One would like a measure that remains sensitive to localized contour mismatch, since small but visible deviations may occur only over a restricted angular sector. At the same time, 
the absolute maximum can be governed by a tiny number of extreme-angle outliers and may therefore exaggerate the typical error. The percentile statistic $\Delta r_{95}$ provides a more stable measure of 
contour-level agreement.

Taken together, the three diagnostics probe different aspects of the same geometric question. The quantity $\Delta A$ tests the fidelity of the integrated shadow area. The quantity $\Delta D_{\rm eq}$ tests 
the preservation of the derived size scale. The quantity $\Delta r_{95}$ tests the agreement of the contour itself, angle by angle, over almost the entire boundary. In this way the reconstruction analysis 
goes beyond the usual practice of characterizing a shadow only through size, circularity, or similar integrated observables \cite{PhysRevD.80.024042,PhysRevD.100.044057,KumarGhosh2020}. Here the object being 
tested is the contour itself, and the diagnostics are chosen precisely to reveal how faithfully that contour survives in the low-order descriptor representation.

\subsection{Exact-level reconstruction results}
\label{subsec:exact_recon_results}

\begin{figure}[t]
\centering
\includegraphics[width=\linewidth]{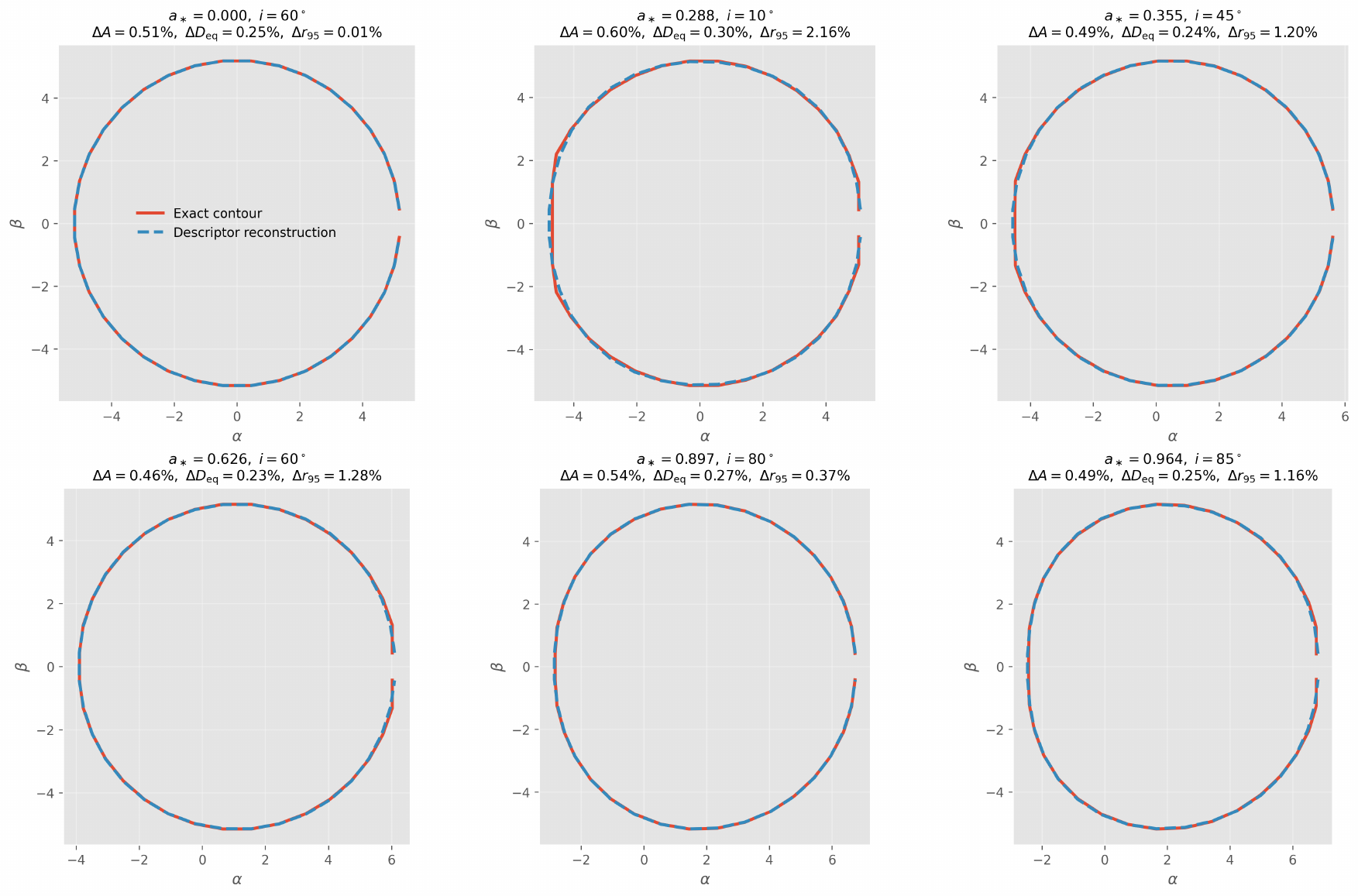}
\caption{
Representative exact descriptor-level reconstructions of Kerr shadow contours
for selected points in the $(a_\ast,i)$ domain.
In each panel, the exact contour is compared directly with the contour reconstructed
from the low-order descriptor set
$\{\bar R,\alpha_c,c_2,c_3,c_4\}$, without any surrogate interpolation.
The reported diagnostics are the relative area error $\Delta A$,
the relative equivalent-diameter error $\Delta D_{\rm eq}$,
and the contour-level radial mismatch $\Delta r_{95}$.
The close overlap shows that the descriptor basis already retains the dominant
global contour geometry at the exact level.
}
\label{fig:exact_recon_examples}
\end{figure}

The descriptor-level reconstruction based on the exact Kerr contour already reproduces the shadow with controlled accuracy throughout the physically relevant parameter domain. Representative examples, 
shown in Fig.~\ref{fig:exact_recon_examples}, make this point quite clearly. The reconstructed contours follow the exact critical curves closely at low, intermediate, and high spin, and for a wide range of 
inclinations. What is important here is that the agreement is not confined to the overall scale alone. The horizontal displacement of the contour and the leading fore--aft asymmetry, which are among the most 
characteristic geometric effects of Kerr rotation, are also retained directly at the contour level.

\begin{figure*}[t]
\centering
\includegraphics[width=\textwidth]{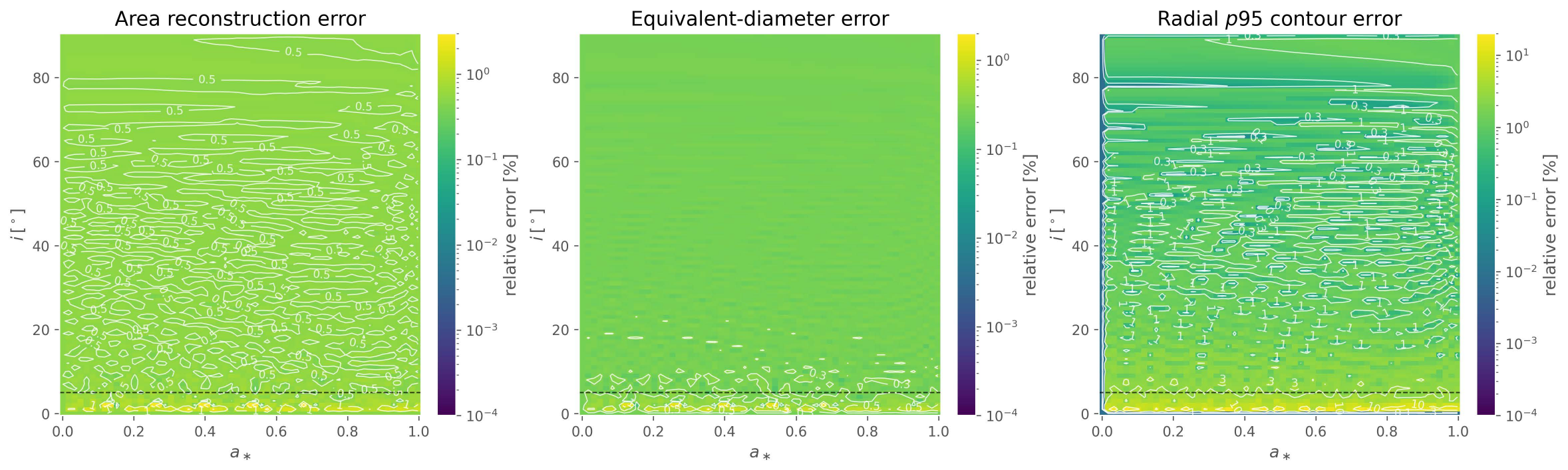}
\caption{
Global exact-level reconstruction errors over the Kerr parameter domain.
Shown are the relative area error $\Delta A$ (left),
the relative equivalent-diameter error $\Delta D_{\rm eq}$ (center),
and the contour-level radial mismatch $\Delta r_{95}$ (right),
computed by comparing the exact Kerr contour with the contour reconstructed
from the exact low-order descriptor set
$\{\bar R,\alpha_c,c_2,c_3,c_4\}$.
The dashed horizontal line marks the bulk-domain threshold $i=5^\circ$.
The dominant degradation is confined to the near-polar strip,
while the bulk of parameter space remains under controlled exact-level truncation error.
}
\label{fig:exact_recon_maps}
\end{figure*}

A more complete picture is provided by Fig.~\ref{fig:exact_recon_maps}, where we display the exact-level error maps over the full $(a_\ast,i)$ domain for the three diagnostics introduced above. The significance 
of these maps is not merely that the errors are numerically small. The more important point is that they remain controlled across the bulk of parameter space even though the contour has been reduced to the 
five-dimensional descriptor set $\mathcal D$. Thus the truncation to
\(
\{\bar R,\alpha_c,c_2,c_3,c_4\}
\)
does not destroy the dominant global geometry of the Kerr shadow. One can say it differently: the essential information carried by the contour is already concentrated in a remarkably small number of low-order 
geometric components.

This conclusion is summarized quantitatively in Table~\ref{tab:exact_recon_summary}. Over the full grid, the exact descriptor reconstruction gives median errors of $0.514\%$ in area, $0.257\%$ in equivalent 
diameter, and $0.937\%$ in the contour-level radial mismatch. If one restricts attention to the bulk domain $i\ge 5^\circ$, these values become $0.513\%$, $0.257\%$, and $0.909\%$, respectively, with no 
negative-radius samples at all. The last observation is worth noting, since it shows that the low-order reconstruction remains geometrically regular over this domain. Taken together, these numbers show that 
the descriptor basis is not merely compact, but already has substantial intrinsic reconstruction power before any surrogate approximation is introduced.

\begin{table}[t]
\centering
\caption{
Summary of exact descriptor-level reconstruction errors for the basis
$\{\bar R,\alpha_c,c_2,c_3,c_4\}$.
}
\label{tab:exact_recon_summary}
\begin{tabular}{lccccccc}
\hline
Regime & $n$
& med.\ $\Delta A$
& p95 $\Delta A$
& med.\ $\Delta D_{\rm eq}$
& p95 $\Delta D_{\rm eq}$
& med.\ $\Delta r_{95}$
& p95 $\Delta r_{95}$ \\
\hline
All nodes & 5460 & 0.514 & 0.611 & 0.257 & 0.306 & 0.937 & 2.963 \\
$i\ge 5^\circ$ & 5160 & 0.513 & 0.571 & 0.257 & 0.286 & 0.909 & 2.508 \\
\hline
\end{tabular}
\end{table}

These exact-level results establish that the descriptor layer is not merely a convenient
regression target. It already functions as a compact geometric representation of the Kerr
shadow in its own right. The later surrogate therefore approximates an object that has
independent geometric meaning, rather than rescuing an otherwise inadequate basis.

\subsection{Role of the exact-level test}
\label{subsec:exact_recon_role}

This exact reconstruction step forms the natural geometric bridge between the full Kerr critical curve and the surrogate model introduced below. Its role is to separate the intrinsic compression error of 
the descriptor basis from any further error associated with interpolation or emulation. This distinction is essential for the interpretation of the later surrogate diagnostics. Once the truncation error at 
the exact level is known, the errors obtained on held-out surrogate points can be read for what they really are: not a failure of the geometric representation itself, but the additional price paid for emulating 
an object that already has independent geometric meaning.

At the same time, the exact-level test clarifies the status of the descriptor basis in a more fundamental way. The set

\(
\{\bar R,\alpha_c,c_2,c_3,c_4\}
\)
is not merely a convenient fitting device. It is a compact low-order basis that already reproduces the dominant geometry of the Kerr contour with controlled error. In this precisely defined sense, the 
descriptor layer carries genuine contour-level information beyond any single size observable. What is being compressed here is not just a number, but the leading geometric content of the shadow boundary itself.


\section{Boundary-aware surrogate construction}
\label{sec:surrogate}

The results obtained above show that the descriptor layer already retains the main geometric information of the Kerr shadow. For this reason, the surrogate problem can now be formulated in a simpler and more 
transparent way. We no longer seek to approximate the full contour directly. Instead, it is enough to interpolate a small set of geometrically meaningful descriptors over the parameter space.

At the same time, this interpolation cannot be treated as a purely numerical task. Near the Schwarzschild and polar limits the shadow approaches a circle, and the nontrivial shape descriptors must vanish in a 
regular way. A reliable surrogate must therefore respect these boundary constraints together with the smooth behavior in the interior of the domain.

\subsection{Parameter domain and training protocol}

We work on a dense rectangular grid in the two physical parameters, namely the dimensionless spin \(a_*\in[0,0.998]\) and the observer inclination \(i\in[0^\circ,90^\circ]\). The full production dataset 
is constructed on a \(60\times 91\) grid. For the held-out interpolation test discussed below, the training set is chosen more sparingly in the spin direction: we retain a \(31\times 91\) subgrid obtained 
by taking every second spin node, while always keeping the endpoint. Validation is then performed on the complementary spin values at each fixed inclination. In this way, the train--validation split probes 
interpolation along the spin axis in a transparent manner, while the sampling in inclination remains explicit throughout.

It follows that the held-out validation used here tests interpolation in the spin direction at fixed sampled inclinations. The reported validation should therefore not be read as a fully general two-dimensional 
leave-out test over both parameters. In the production setting considered here, the inclination grid is kept explicit, while the regularized descriptor fields are interpolated in the spin variable at each 
inclination node. For the contour-scan problem studied in the present work, this construction is sufficient and natural; a genuinely two-dimensional interpolation and validation protocol would be a natural 
extension.

For clarity, it is useful to summarize the computational workflow in the order in which it is actually used, and this summary is collected in Table~\ref{tab:workflow_summary}. The exact contour generation and 
descriptor preparation are performed offline, whereas the surrogate is evaluated only at the level of the regularized descriptor fields and is then mapped back to the contour.

\begin{table}[t]
\centering
\caption{
Algorithmic workflow of the contour-level surrogate, with the separation between offline preparation and online evaluation made explicit.
}
\label{tab:workflow_summary}
\footnotesize
\renewcommand{\arraystretch}{1.08}
\begin{tabularx}{\linewidth}{cXc}
\hline
Stage & Operation & Mode \\
\hline
1 & Compute the exact Kerr contour from the spherical-photon critical curve on the production grid. & Offline \\
2 & Center the contour and resample the radial profile on the common angular grid used throughout the paper. & Offline \\
3 & Extract the descriptor set \(\bar R,\alpha_c,c_2,c_3,c_4\). & Offline \\
4 & Transform these descriptors to the regularized fields \(G_0,G_1,U_2,U_3,U_4\), impose the exact boundary values, and store the resulting grid data. & Offline \\
5 & At a requested \((a_*,i)\), interpolate the regularized fields in the spin variable at the sampled inclination node by monotone piecewise-cubic Hermite interpolation. & Online \\
6 & Invert the transforms and reconstruct the centered radial contour profile \(R(\psi)\). & Online \\
7 & Apply the positivity safeguard if needed and recover the final contour and derived observables. & Online \\
\hline
\end{tabularx}
\end{table}

Unless stated otherwise, the numerical settings collected here are those of the distributed production run. The full parameter grid contains \(60\times 91\) nodes in \((a_*,i)\), while the held-out validation is 
based on the \(31\times 91\) spin-thinned training subgrid described above. In the accompanying offline profile-building script, the exact contour generation uses \(n_r=300\) samples along the spherical-photon 
branch, and circular limiting cases are represented by polygons with \(360\) points. The saved exact radial profiles used by the surrogate are represented on the uniform angular grid introduced earlier with 
\(N_\psi=36\) samples and are stored in standard double precision (\texttt{float64}). At each sampled inclination, the regularized descriptor fields are interpolated in the spin variable by the SciPy 
implementation \texttt{PchipInterpolator}. The regularization floor is \(\varepsilon=10^{-6}\), and the positivity safeguard is applied with \(S_{\max}=0.49\).

Since the practical role of a surrogate also depends on its computational cost, it is useful to record the measured timing of the present implementation, and this summary is collected in Table~\ref{tab:cost_summary}. 
The timings reported there were obtained from \(10^4\) held-out bulk evaluations with \(i\ge 5^\circ\), using the same production settings just described.

\begin{table}[t]
\centering
\caption{
Measured computational cost of the exact contour pipeline and of the descriptor surrogate in the present implementation. The online timings are reported per contour from \(10^4\) held-out bulk evaluations.
}
\label{tab:cost_summary}
\footnotesize
\renewcommand{\arraystretch}{1.08}
\begin{tabularx}{\linewidth}{>{\raggedright\arraybackslash}p{0.25\linewidth}XX}
\hline
Quantity & Exact contour pipeline & Descriptor surrogate \\
\hline
Offline preparation & None for a single exact evaluation; a tabulated \(60\times 91\) reference grid with \(N_\psi=36\) sampled profiles takes \(210.39\,\mathrm{s}\). & The same exact reference build plus the descriptor-table fit; the additional fit takes \(0.059\,\mathrm{s}\), for \(210.45\,\mathrm{s}\) in total. \\
Stored contour representation per grid point & One horizontal centroid shift together with a sampled radial profile of \(N_\psi=36\) radii. & Five real descriptor values, \(\bar R,\alpha_c,c_2,c_3,c_4\). \\
Online evaluation & Fresh exact critical-curve sampling with \(n_r=300\). & Five one-dimensional monotone piecewise-cubic Hermite interpolations, inverse transforms, and harmonic contour reconstruction. \\
Median runtime per contour & \(40.73\,\mathrm{ms}\) \(\bigl(43.76\,\mathrm{ms}\ \text{at}\ p95\bigr)\). & \(0.0257\,\mathrm{ms}\) \(\bigl(0.0362\,\mathrm{ms}\ \text{at}\ p95\bigr)\). \\
Speed-up & --- & \(1.58\times 10^3\) by the ratio of median runtimes. \\
\hline
\end{tabularx}
\end{table}

One can thus see that repeated contour scans do become substantially faster at the surrogate stage. At the same time, the central point of the present construction remains the compact descriptor representation itself, 
while the acceleration is a useful consequence rather than the sole motivation.

To capture the expected behaviour near the boundaries of parameter space, it is convenient to introduce the auxiliary variables
\begin{equation}
x \equiv a_*^2,
\qquad
z \equiv \sin^2 i,
\label{eq:xz_def}
\end{equation}
for which both the Schwarzschild limit \(a_*\to 0\) and the polar limit \(i\to 0\) correspond to the same boundary \(xz\to 0\). This choice is not merely a change of notation. It reflects the fact that the 
leading structure of the descriptor fields near these limits is naturally organized in terms of even powers of the spin and of the inclination through \(\sin i\). The variables \(x\) and \(z\) therefore provide 
a more suitable language for imposing the correct regularity of the surrogate near the edges of the physical domain.

We also make use of the exact polar shadow radius introduced in Sec.~\ref{sec:exact_pipeline}, writing it in the normalized form
\begin{equation}
y_{\rm pole}(a_*) \equiv \frac{\rho_{\rm pole}(a_*)}{3\sqrt{3}},
\qquad
\rho_{\rm pole}(a_*) = 3\sqrt{3}\,y_{\rm pole}(a_*).
\label{eq:ypole_def}
\end{equation}
The normalization by \(3\sqrt{3}\) is natural, since this is the Schwarzschild shadow radius in the units used here. Thus \(y_{\rm pole}(a_*)\) measures the polar shadow radius relative to the nonrotating 
reference value, and it provides a convenient exact boundary quantity for the construction that follows.

\subsection{Transformed variables and regularized descriptor fields}

The descriptor fields
\(
\bar R(a_*,i),\alpha_c(a_*,i),c_2(a_*,i),c_3(a_*,i),c_4(a_*,i)
\)
develop a pronounced boundary structure as one approaches the Schwarzschild limit or the polar limit. It is therefore convenient, before performing any interpolation, to remove this leading boundary 
behaviour explicitly. The idea is simple: one factors out the known scaling near the boundaries so that the remaining fields become smoother and numerically better behaved.

To this end, we introduce the auxiliary quantities
\begin{equation}
\varepsilon \equiv 10^{-6},
\qquad
d_{xz} \equiv \max(xz,\varepsilon),
\qquad
d_a \equiv \max(\sqrt{xz},\sqrt{\varepsilon}).
\label{eq:denom_defs}
\end{equation}
The same floor \(d_{xz}\) is used for both the mean-radius and harmonic sectors, so no separate notation is needed there. These denominators are chosen so that the transformed variables remain finite and 
regular even when \(a_*\to 0\) or \(i\to 0\). In this way, the singular-looking behaviour belongs only to the explicit prefactors, while the quantities to be interpolated remain smooth.

We then define the transformed fields by
\begin{equation}
G_0
\equiv
-\frac{1}{d_{xz}}
\ln\!\left(\frac{\bar R}{3\sqrt{3}\,y_{\rm pole}(a_*)}\right),
\label{eq:G0_def}
\end{equation}
\begin{equation}
G_1
\equiv
\frac{\alpha_c}{d_a},
\label{eq:G1_def}
\end{equation}
and
\begin{equation}
U_2
\equiv
\operatorname{arsinh}\!\left(\frac{c_2}{d_{xz}}\right),
\qquad
U_3
\equiv
\operatorname{arsinh}\!\left(\frac{c_3}{d_{xz}}\right),
\qquad
U_4
\equiv
\operatorname{arsinh}\!\left(\frac{c_4}{d_{xz}}\right).
\label{eq:Un_def}
\end{equation}
The meaning of these definitions is transparent. The quantity \(G_0\) measures the deviation of the mean radius from the exact polar reference value after the leading boundary suppression has been 
removed. The field \(G_1\) does the same for the horizontal centroid shift. The variables \(U_2,U_3,U_4\) play an analogous role for the harmonic coefficients, with the inverse hyperbolic sine chosen 
to keep the transformed fields well scaled over the full domain without losing sensitivity to small values.

On the exact Schwarzschild or axial boundaries, the nontrivial contour structure disappears altogether. The shadow becomes circular, the centroid shift vanishes, and the harmonic deformations are 
absent. It is therefore natural to impose
\begin{equation}
G_0=G_1=U_2=U_3=U_4=0.
\label{eq:boundary_zeroing}
\end{equation}
This is not an additional model assumption, but simply the direct expression of the known boundary geometry in the transformed variables.

At each fixed inclination \(i\), we then interpolate the transformed fields as functions of the spin parameter \(a_*\) by monotone piecewise-cubic Hermite interpolation. Thus,
\begin{equation}
G_\mu(a_*,i)\approx \mathcal{I}^{\rm PCHIP}_i[G_\mu](a_*),
\qquad
\mu=0,1,
\end{equation}
and
\begin{equation}
U_n(a_*,i)\approx \mathcal{I}^{\rm PCHIP}_i[U_n](a_*),
\qquad
n=2,3,4.
\end{equation}
This choice gives a numerically stable production model for the descriptor layer while preserving the physical meaning of the descriptors after the transformation is inverted. In other words, 
the interpolation is carried out not on arbitrary fit parameters, but on regularized geometric fields whose limiting behaviour is already built in.

Once the interpolated transformed variables
\(
G_{0,{\rm pred}},G_{1,{\rm pred}},U_{2,{\rm pred}},U_{3,{\rm pred}},U_{4,{\rm pred}}
\)
have been obtained, the physical descriptors are reconstructed by inverting the transforms:
\begin{equation}
\bar R_{\rm pred}
=
3\sqrt{3}\,y_{\rm pole}(a_*)
\exp\!\bigl(-d_{xz}\,G_{0,{\rm pred}}\bigr),
\label{eq:Rbar_pred}
\end{equation}
\begin{equation}
\alpha_{c,{\rm pred}}
=
d_a\,G_{1,{\rm pred}},
\label{eq:alpha_pred}
\end{equation}
and
\begin{equation}
c_{2,{\rm pred}}
=
d_{xz}\,\sinh(U_{2,{\rm pred}}),
\qquad
c_{3,{\rm pred}}
=
d_{xz}\,\sinh(U_{3,{\rm pred}}),
\qquad
c_{4,{\rm pred}}
=
d_{xz}\,\sinh(U_{4,{\rm pred}}).
\label{eq:cn_pred}
\end{equation}
Thus the regularized interpolation problem is brought back to the original geometric quantities in a completely explicit way.

Finally, on the exact boundaries \(a_*=0\) or \(i=0\), we impose the limiting behaviour directly:
\begin{equation}
\alpha_{c,{\rm pred}}=0,
\qquad
c_{2,{\rm pred}}=c_{3,{\rm pred}}=c_{4,{\rm pred}}=0,
\qquad
\bar R_{\rm pred}=3\sqrt{3}\,y_{\rm pole}(a_*).
\label{eq:boundary_physical_descriptors}
\end{equation}
This closes the construction in the natural way. The surrogate is therefore boundary-aware from the outset: it respects the exact circular limits where the contour loses its nontrivial 
structure, while away from those limits it provides a smooth interpolation of the descriptor fields that still retain their direct geometric interpretation.

\subsection{Positivity safeguard and profile reconstruction}

The predicted descriptors determine the surrogate radial profile in a direct way:
\begin{equation}
R_{\rm pred}(\psi)
=
\bar R_{\rm pred}
\Bigl[
1+2\bigl(
c_{2,{\rm pred}}\cos 2\psi
+
c_{3,{\rm pred}}\cos 3\psi
+
c_{4,{\rm pred}}\cos 4\psi
\bigr)
\Bigr].
\label{eq:Rpred_raw}
\end{equation}
At this point one must ensure that the reconstructed profile remains physically meaningful. Since \(R_{\rm pred}(\psi)\) represents the distance from the centroid to the contour, it must stay positive 
for every angle \(\psi\). A negative value would have no geometric interpretation and would signal that the harmonic correction has become too large.

To control this, we introduce the total harmonic amplitude
\begin{equation}
S_{\rm harm}
\equiv
|c_{2,{\rm pred}}|+|c_{3,{\rm pred}}|+|c_{4,{\rm pred}}|.
\label{eq:Sharm_def}
\end{equation}
The usefulness of this quantity is immediate, because one always has
\begin{equation}
\left|
c_{2,{\rm pred}}\cos 2\psi
+
c_{3,{\rm pred}}\cos 3\psi
+
c_{4,{\rm pred}}\cos 4\psi
\right|
\le
S_{\rm harm}.
\end{equation}
It follows that a sufficient condition for the positivity of \(R_{\rm pred}(\psi)\) for all \(\psi\) is
\begin{equation}
2S_{\rm harm}<1.
\label{eq:positivity_condition}
\end{equation}
This condition is not sharp, but it is simple, transparent, and entirely adequate for the present purpose. It guarantees that the harmonic deformation never overwhelms the mean radius.

In the production model we enforce this requirement by a minimal global rescaling of the harmonic sector. We define the shrink factor
\begin{equation}
\lambda
\equiv
\min\!\left(1,\frac{S_{\max}}{S_{\rm harm}}\right),
\qquad
S_{\max}=0.49,
\label{eq:lambda_def}
\end{equation}
and replace the predicted coefficients by
\begin{equation}
\tilde c_n = \lambda\, c_{n,{\rm pred}},
\qquad
n=2,3,4.
\label{eq:cn_shrunk}
\end{equation}
Thus nothing is changed when the harmonic amplitudes are already safely inside the allowed region. Only when the positivity bound is approached do we reduce them, and then only by the smallest 
factor needed to restore a margin of safety.

The final safeguarded profile therefore takes the form
\begin{equation}
R_{\rm pred}(\psi)
=
\bar R_{\rm pred}
\Bigl[
1+2\bigl(
\tilde c_2\cos 2\psi
+
\tilde c_3\cos 3\psi
+
\tilde c_4\cos 4\psi
\bigr)
\Bigr].
\label{eq:Rpred_final}
\end{equation}
The corresponding reconstructed contour on the observer screen is
\begin{equation}
\alpha_{\rm pred}(\psi)=\alpha_{c,{\rm pred}}+R_{\rm pred}(\psi)\cos\psi,
\qquad
\beta_{\rm pred}(\psi)=R_{\rm pred}(\psi)\sin\psi,
\label{eq:predicted_contour}
\end{equation}
where, in the present centered convention, \(\beta_c=0\).

It is worth emphasizing that this safeguard is deliberately modest. It does not modify the descriptor structure unless the predicted harmonic content comes close to violating positivity. In other words, 
it is not an additional model layer, but only a protection against an obviously unphysical outcome. Over the bulk-domain production evaluation, this simple procedure yields zero negative-radius samples. 
In the held-out bulk set \(i\ge 5^\circ\), the safeguard is in fact never activated: the condition \(\lambda<1\), equivalently \(S_{\rm harm}>S_{\max}\), occurs for \(N_{\rm trig}=0\) out of 
\(N_{\rm bulk}=2494\) samples, so \(f_{\rm trig}=100N_{\rm trig}/N_{\rm bulk}=0.000\%\). Correspondingly, the median and minimum shrink factors are both \(\lambda=1\), the maximum shrink \(1-\lambda\) is zero, 
and the largest raw value of \(S_{\rm harm}\) on this subset is \(0.0777\), well below \(S_{\max}=0.49\). Thus the positivity layer acts as a robustness safeguard rather than as a systematic deformation of the 
bulk-domain surrogate, and there is no rescaling-induced contour bias on the reported held-out bulk evaluation.
Accordingly, the surrogate remains not only compact and inexpensive, but also geometrically well defined throughout the region of interest.


\section{Accuracy diagnostics in the bulk domain}
\label{sec:accuracy}

The main validation results are collected in Table~\ref{tab:production_metrics}, while Fig.~\ref{fig:deq_shape} illustrates the qualitative shape-discrimination example. Our main quantitative claim is 
restricted to the bulk domain

\begin{equation}
i \ge 5^\circ,
\label{eq:bulk_domain}
\end{equation}
where the production surrogate is numerically stable and the descriptor fields vary smoothly.

Within this regime, the agreement between the exact contours and the surrogate reconstructions remains uniformly good. This point is essential. The surrogate is not merely reproducing an integrated size 
scale; it retains the dominant global geometry of the Kerr shadow at the contour level. In most of the bulk domain the errors stay small, and they grow appreciably only as one approaches the near-polar boundary. 
This behaviour is physically natural. Near the axis, the shadow becomes increasingly circular, and the descriptor fields correspondingly become more delicate to interpolate in a robust way.

A global view of the exact descriptor-level reconstruction error is provided by Fig.~\ref{fig:exact_recon_maps}, which shows the corresponding error maps for the three diagnostics introduced in Sec.~\ref{sec:exact_reconstruction}: the relative 
area error \(\Delta A\), the relative equivalent-diameter error \(\Delta D_{\rm eq}\), and the contour-level radial statistic \(\Delta r_{95}\). These maps show that the errors remain low over most of the domain, 
with the same systematic increase only near the polar edge. Thus the loss of accuracy is not widespread, but concentrated precisely where the geometry becomes most nearly degenerate.

The numerical summary in Table~\ref{tab:production_metrics} makes this more concrete. Over the held-out bulk domain, the production surrogate achieves median errors of \(0.522\%\) in area, \(0.261\%\) in 
equivalent diameter, and \(0.954\%\) in the radial-\(p95\) contour reconstruction statistic. The corresponding \(p95\) values are \(1.100\%\), \(0.551\%\), and \(3.436\%\). In the same regime, the safeguarded 
reconstruction yields no negative-radius samples. This last point is worth noting, since it shows that the surrogate remains geometrically well behaved throughout the bulk domain.

Taken together, these results show that the production model provides a controlled contour-level approximation across the physically relevant bulk of parameter space. Compact surrogates of this kind are useful 
whenever the contour must be evaluated repeatedly, for example in parameter scans and contour-level comparisons \cite{Tiede2022_ngEHT_PhotonRing,Walia_2025,EHTC2025_M87_Persistent_II}. More importantly for the 
present paper, they show that the low-order descriptor layer remains geometrically useful even after surrogate emulation. The compact representation therefore stays accurate not only for integrated observables, 
but for the contour itself.

\begin{table}[t]
\centering
\caption{
Accuracy summary for the production contour surrogate.
The primary claim of the paper is quoted for the bulk held-out domain \(i\ge 5^\circ\).
}
\scriptsize
\setlength{\tabcolsep}{3.5pt}
\renewcommand{\arraystretch}{1.05}
\begin{tabular}{lrrrrrrrr}
\hline
Regime & \(n\) & med.\(A\) & p95\(A\) & med.\(D_{\rm eq}\) & p95\(D_{\rm eq}\) & med.\(r_{95}\) & p95\(r_{95}\) & neg. \\
& & [\%] & [\%] & [\%] & [\%] & [\%] & [\%] & radii \\
\hline
Bulk held-out (\(i\ge 5^\circ\)) & 2494 & 0.522 & 1.100 & 0.261 & 0.551 & 0.954 & 3.436 & 0 \\
\hline
\end{tabular}
\label{tab:production_metrics}
\end{table}


\section{Information content beyond equivalent diameter}
\label{sec:information}

A basic limitation of any size-only description is immediately clear: a full contour is reduced to a single scalar. Such reductions are standard and often useful in the shadow literature, where one commonly 
characterizes the image by a typical size, a circularity measure, a distortion parameter, an asymmetry index, or some related integrated observable \cite{PhysRevD.80.024042,PhysRevD.100.044057,KumarGhosh2020}. 
But for the Kerr shadow the weakness of this reduction can be seen explicitly. Nearly equal equivalent diameter does not imply nearly equal contour geometry. The same general difficulty appears more broadly in 
studies of non-Kerr or modified shadow families, where a small set of global observables often provides only the first coarse characterization 
\cite{Sui2023_Accel_KerrNewman_ArealRadius,AgurtoSepulveda2024_AlphaCorrectedShadows,Kuang:2024ugn,Ban2025_EPJC_QGShadows,PhysRevD.108.044008}.

This is illustrated directly in Fig.~\ref{fig:deq_shape}. Each row there shows a pair of Kerr shadows with almost the same value of \(D_{\rm eq}\), yet with clearly different contour shapes. If one 
retains only \(D_{\rm eq}\), then both members of the pair are mapped to nearly the same diameter-equivalent circle and become, for practical purposes, indistinguishable at the contour level. 
This example is not intended as a parameter-inference validation. The low-order descriptor reconstruction behaves differently: it distinguishes the two shapes and reproduces their differences in close agreement 
with the exact Kerr contours. The point is simpler: it serves as a diagnostic illustration showing why a single size observable is not by itself an adequate contour representation.

It follows that the descriptor layer contains geometric information that is absent from any single size observable. In particular, the centroid 
shift \(\alpha_c\) and the leading low-order harmonic coefficients \((c_2,c_3,c_4)\) retain aspects of the contour that are necessarily lost when the shadow is compressed to one number. The descriptor set 
should therefore be viewed not only as a compact parametrization, but as an intermediate representation of the Kerr shadow with a direct geometric meaning.

In this sense, the role of the surrogate is broader than that of a fast approximation for a single observable. Its purpose is to emulate a low-dimensional and physically interpretable representation of the 
shadow contour itself.

\section{Discussion and limitations}
\label{sec:limitations}

The results obtained above support a geometric interpretation of the surrogate that is more informative than any approximation based on size alone. The essential point 
is not merely that the production model reproduces the contour with low held-out error over the bulk domain, but that the descriptor layer on which it is built already has 
a direct physical meaning. The mean radius \(\bar R\), the centroid shift \(\alpha_c\), and the low-order harmonic coefficients \((c_2,c_3,c_4)\) together provide a 
compact description of the dominant global features of the Kerr shadow: its overall scale, its horizontal displacement, and its leading departures from circularity. In 
this sense, the surrogate acts not on an opaque latent representation, but on a low-dimensional contour geometry whose components can be interpreted directly.

This viewpoint also makes clear the role of the model within the present work. Horizon-scale observations and model-comparison studies already motivate the search for 
compact geometric summaries of black-hole shadow structure, while future programs with higher angular resolution are likely to increase the usefulness of fast contour-level 
representations 
\cite{EHTC2024_M87_Persistent_I,EHTC2025_M87_Persistent_II,EHTC2024_SgrA_Polarization_VII,EHTC2024_SgrA_Polarization_VIII,Tiede2022_ngEHT_PhotonRing,Walia_2025,Wielgus2024_SgrA_InternalFaraday,Vagnozzi:2023SgrA_CQG,Khodadi:2024MimeticSR}. 
Because the descriptor set retains contour-level information beyond the equivalent diameter alone, it may serve naturally as an intermediate geometric layer in rapid 
parameter scans and contour-based comparisons, where one wishes to preserve more information than a single scalar size measure without passing 
immediately to full image-domain modelling. What the present construction shows is that a small 
set of interpretable global descriptors already captures a nontrivial part of the Kerr-shadow geometry.

At the same time, the scope of the construction is deliberately limited. A first restriction concerns the near-polar strip \(i<5^\circ\), which remains numerically delicate 
in the present production model. In this regime the Kerr shadow approaches a circle, so the genuinely shape-carrying variations become strongly compressed, while relative 
contour-level error measures become increasingly sensitive even to very small absolute mismatches. The same limit is also more difficult from the interpolation point of view, 
since the nontrivial descriptor amplitudes are driven toward their boundary values. This limitation is therefore primarily algorithmic rather than fundamental. The descriptor 
basis itself does not become ambiguous in the near-polar regime: the exact centered radial representation remains well defined, and the exact descriptor-level reconstruction 
stays controlled over the full domain. What becomes delicate is the surrogate stage, because the noncircular descriptor amplitudes are driven toward zero near the polar 
boundary, while relative contour-level diagnostics become increasingly sensitive to very small absolute mismatches. For this reason, the main quantitative claim of the paper is restricted to the bulk 
domain \(i\ge 5^\circ\), where the descriptor fields vary smoothly and the held-out errors remain well controlled.

A second restriction is conceptual. The model developed here is not intended to replace full ray tracing, radiative-transfer calculations, instrumental response modelling, 
or image-domain forward pipelines. These belong to a broader observational and theory-comparison framework than the one considered in the present paper 
\cite{EHTC2024_M87_Persistent_I,EHTC2025_M87_Persistent_II,Vagnozzi:2023SgrA_CQG,Khodadi:2024MimeticSR}. More specifically, the surrogate constructed here represents the pure Kerr critical curve itself, without 
astrophysical blurring, radiative-transfer emission structure, instrumental convolution, or image-reconstruction effects. Our aim is different. What is constructed here is a compact 
contour-level geometric module derived directly from the exact Kerr critical curves. Such a module is naturally useful for geometric compression, rapid parameter surveys, 
and contour-level benchmarks, but it is not a substitute for a complete observational pipeline.

These limitations should therefore be understood as part of the intended scope of the work rather than as a weakness of its main result. The central claim is simply that 
the Kerr shadow admits a compact and physically interpretable low-order contour representation, and that this representation can be both reconstructed and emulated with 
controlled accuracy over the physically relevant bulk domain.

\section{Conclusion}
\label{sec:conclusion}

We have constructed a compact low-order representation of the Kerr shadow contour by extracting a small set of descriptors directly from the exact critical curve. 
The construction is based on quantities with a clear geometric meaning: the mean radius \(\bar R\), the horizontal centroid shift \(\alpha_c\), and the leading harmonic 
coefficients \((c_2,c_3,c_4)\). Together they describe the dominant global features of the contour, namely its overall scale, its displacement on the observer screen, and 
its leading deviations from circularity.

The first result of the paper is that this descriptor set is already nontrivial at the level of exact geometry. When it is extracted from the exact Kerr contour, it 
reproduces the dominant contour shape with controlled errors in the enclosed area, in the equivalent diameter, and in the radial contour mismatch. This shows that the 
descriptor basis is not merely a convenient target for regression. It is already a genuine compression layer for Kerr-shadow geometry.

The second result is that these descriptors can be emulated accurately across parameter space by a boundary-aware surrogate. Over the held-out bulk domain \(i\ge 5^\circ\), 
the production model achieves median errors of \(0.522\%\) in area, \(0.261\%\) in equivalent diameter, and \(0.954\%\) in the radial-\(p95\) contour reconstruction 
statistic, with corresponding \(p95\) values of \(1.100\%\), \(0.551\%\), and \(3.436\%\). At the same time, the safeguarded reconstruction produces no negative-radius 
samples. This shows that the compact descriptor layer remains stable and geometrically useful even after surrogate emulation.

The third result is conceptual. Nearly equal equivalent diameter does not imply the same Kerr shadow shape. The low-order descriptor representation retains contour-level 
distinctions that are necessarily lost when the shadow is reduced to a single size observable. In this sense, the surrogate does more than approximate one number. It 
emulates a compact and physically meaningful representation of the shadow contour itself.

Taken together, these results show that the Kerr shadow admits an intermediate description that lies between a single scalar observable and the full exact contour. This 
is precisely what gives the present construction its usefulness. On the one hand, it provides a practical contour-level surrogate. On the other hand, it provides a compact 
geometric language for the shadow itself. Natural extensions of the present work include a more robust treatment of the near-polar regime, the extension of the descriptor 
framework to non-Kerr shadow families, and the formulation of future degeneracy studies at the descriptor level.

\printbibliography[title={References}]

\end{document}